\documentclass{agujournal2019}
\usepackage{lmodern}
\usepackage{url} 
\usepackage{multirow}
\usepackage[inline]{trackchanges} 
\usepackage{soul}
\usepackage{float} 
\justifying

\draftfalse

\journalname{The Journal of Advances in Modeling Earth Systems}

\begin{document}

%
%


  \title{Optimizing Objective Model Calibration Approaches using Single Column Models}

%
%




\authors{Pappu Paul\affil{1}, Cristian Proistosescu,\affil{1,2}}


\affiliation{1}{Department of Climate, Meteorology \& Atmospheric Sciences, University of Illinois Urbana Champaign}
\affiliation{2}{Department of Earth Sciences \& Environmental Change, University of Illinois Urbana Champaign}




\correspondingauthor{Pappu Paul}{pappup2@illinois.edu}



\begin{keypoints}
\item Single Column Models are an efficient tools for calibrating complex Earth system models.
\item Observation-fitting calibration can appear successful, but may achieve good agreement for the wrong reasons.
\item Bayesian approach shows strong promise for constraining parameter posterior distributions.
\end{keypoints}

%
%

%
%


\begin{abstract}
Sub-grid scale parameterizations in atmospheric models involve numerous uncertain parameters that must be tuned to align simulations with observations. Here, we propose a framework for assessing objective tuning frameworks using the Single Column Atmosphere Model (SCAM), which retains key physical parameterizations of general circulation models (GCMs) while greatly reducing computational cost. We conduct a perfect-model experiment where we run SCAM with a known “true” parameter set to generate synthetic observations that mimic Atmospheric Radiation Measurement (ARM) Intensive Observation Periods. Perturbed parameter ensembles are constructed by varying microphysics, convection, and aerosol parameters, and cloud–radiation fields are evaluated over the Southern Great Plains. We find that point estimates find solutions that greatly reduce model–observation misfit without recovering the true parameter values. In contrast, a Bayesian framework using a Gaussian Process emulator with Markov Chain Monte Carlo sampling yields tighter constraints on some parameters and more consistent recovery across experiments and variables. The perfect model framework allows to assess which observables yield most information, which parameters are recoverable given a certain set of observations,  and what is the minimum observational record needed. Although this study focuses on a single location with synthetic observations, such experiments provide a controlled setting to evaluate and identify robust calibration frameworks, which can then be extended to multiple locations and real observations with greater confidence. 

\end{abstract}

\section*{Plain Language Summary}
This study develops systematic ways to tune uncertain parameters in atmospheric models using a simplified model. We generate artificial observations from a known model state and test whether different methods can recover the original parameters. We find that simply choosing the best-fitting parameter set is not reliable, and increasing data or sample size does not always improve results. In contrast, a probabilistic approach provides more consistent and accurate identification of the most important parameters, especially with 60-day simulation lengths and larger samples. Although we test an idealized case with artificial data, the method can be extended to real observations and more complex models, offering a more efficient way to reduce uncertainty in climate model tuning.
%
%

%


%
%
%
%

\section{Introduction}
\subsection{The Model Calibration Problem}
General Circulation Models (GCMs) are essential tools for advancing our understanding of Earth system sciences and projecting future atmospheric changes. However, due to computational limitations, GCMs are unable to directly resolve many critical small-scale processes such as microphysics, turbulence, convection, and aerosol interactions. GCMs approximate these unresolved processes using parameterizations: simplified representations with tunable parameters. For example, the parameter `$micro\_mg\_accre\_enhan\_fact$' in the Community Earth System Model (CESM) governs the enhancement factor for raindrop collection of cloud water in the microphysics scheme, directly influencing precipitation formation \cite{gettelman2015advanced, gettelman_mori2015advanced}.

These parameterizations introduce two different types of uncertainty: structural and parametric. Structural uncertainty arises because different parameterization schemes yield different representations of sub-scale processes, all of which are imperfect representations of reality. Structural uncertainty has traditionally been considered the largest source of uncertainty \cite{duffy_perturbing_2023}. 
Parametric uncertainty, on the other hand, refers to uncertainty in the values assigned to the model parameters, with recent studies showing that it can be just as important as structural uncertainty \cite{dunbar_calibration_2021, duffy_perturbing_2023, eidhammer_extensible_2024}. This study focuses only on parametric uncertainty of microphysics, convection and aerosol, where  we want to implement and assess a systematic model tuning (calibration) approach. \par

The calibration process involves optimizing model parameters to improve agreement between simulations and observations while maintaining a fixed model configuration \cite{hourdin_art_2017, schmidt_practice_2017, schneider_earth_2017}. Calibration is an integral part of model development but has often been carried out in an ad hoc and inefficient manner, typically relying on manual adjustments often called ``\textit{hand tuning}" by the model developer community \cite{annan2007efficient, jarvinen2010estimation, gregoire2011optimal,schmidt_practice_2017}. This hand tuning heavily relies on expert judgment and tests only a limited set of parameter set combinations, increasing the risk of overlooking better-performing combinations \cite{gregoire2011optimal}. In addition, these combinations need numerous trial-and-error GCM simulations, making the process slow and computationally expensive \cite{annan2007efficient, jarvinen2010estimation}. Importantly, hand-tuning provides little to no insight into parameter sensitivity or uncertainty, factors that are critical to understanding model behavior and improving predictive reliability.\par 

Given these limitations of traditional hand tuning, there is a clear need for objective model tuning, which can be defined as a systematic, quantitative, and computationally efficient framework for exploring and constraining model parameters. Unlike heuristic approaches, objective methods search the parameter space using formal statistical or optimization techniques, evaluate model–observation misfit using clearly defined metrics, and provide a principled way to quantify uncertainty in parameter estimates. By assessing a broad range of parameter values across multiple variables, regimes, and locations, objective tuning reduces reliance on subjective judgment and offers a more transparent and scientifically defensible foundation for improving climate model performance \cite{annan2007efficient, jarvinen2010estimation, gregoire2011optimal, elsaesser_using_2024, eidhammer_extensible_2024}. \par

\subsection{Ensemble Methods}

Recent efforts to objectively calibrate model parameters have focused on ensemble methods. Most of them are based on a large number of parameter ensemble called Perturbed Parameter Ensembles (PPEs) \cite{ gregoire2011optimal, qian_parametric_2018, eidhammer_extensible_2024, elsaesser_using_2024, yarger_autocalibration_2024} and a few are based on the Ensemble Kalman Filter method \cite<EnKF,>{annan2007efficient, massonnet_calibration_2014, dunbar_calibration_2021, cleary_calibrate_2021}.  \par

To bypass the computational cost of a full numerical calibration, PPE calibration approaches leverage machine learning emulators, trained on PPE ensembles to learn the parameter–climate response relationships. PPE $+$ emulation requires fewer simulations than applying an optimization technique directly to the GCM, but introduces additional uncertainty, related to the emulator's generalizability outside of the training set. EnKF methods, on the other hand, avoid emulator error by directly using GCM simulations, and running the full model with an ensemble of parameter sets over many short assimilation windows. In each iteration observational constraints are assimilated using the Kalman gain to update parameter estimates systematically, which enables convergence toward an improved and dynamically consistent parameter set \cite{evensen_ensemble_2003, massonnet_calibration_2014, cleary_calibrate_2021,dunbar_calibration_2021, sueki_precision_2022,higdon_computer_2012}. 

Still, both PPE and data-assimilation ensemble methods retain significant computational cost. For instance, \citeA{eidhammer_extensible_2024} considers 263-member PPE, each of which are run for three years using three scenario: pre-industrial, present day and future warming, in total of \textit{$263\times3\times3=2367$} years of GCM simulations. Then they use these outcomes to train machine learning (ML) emulators and tune parameters. Another ensemble method, known as the Calibrated Physics Ensemble~\cite<CPE,>{elsaesser_using_2024} begins with a broad range of parameter values (450 ensemble members) and evaluates each ensemble member using a cost function. Only the ensembles that fall within a certain uncertainty range using Markov Chain Monte Carlo (MCMC) sampling, constrained by observations, are selected and used in a more refined calibration process with the aid of ML emulators. Due to their iterative nature EnKF methods are even more computationally expensive, and they also do not explicitly quantify uncertainty. Thus, computational limitations mean that understanding and optimizing these calibration frameworks remains a challenging task.\par

Another fundamental and often overlooked challenge in model calibration is the problem of equifinality. The situation where multiple combinations of parameters produce similarly good agreement with observations \cite{munoz2014identifiability, khatami2019equifinality, whelan2019uncertainty}. This means that even if a model reproduces key observational metrics, the underlying parameter values may not be uniquely constrained. As a result, different parameter sets can yield comparable global statistics (e.g., radiation fields, temperature) while representing very different physical processes or producing divergent future projections. This ambiguity reduces confidence in the physical interpretability and predictive skill of calibrated parameters. 

To address equifinality, calibration frameworks are moving beyond single “best-fit” solutions and instead aim to quantify the full range of plausible parameter combinations and their associated uncertainties. Bayesian probabilistic approaches, particularly MCMC methods like Hamiltonian Monte Carlo (HMC), offer a rigorous pathway to do so by estimating the posterior probability distribution of parameters rather than a single optimum \cite{annan2007efficient, neal2011mcmc, gregoire2011optimal, jarvinen2010estimation, cleary_calibrate_2021, elsaesser_using_2024}. These approaches help distinguish influential parameters, detect compensating errors, and assess the robustness of parameter solutions across different regions and time periods. The computational expense of MCMC methods however, means that they can only be implemented using emulators, not full GCMs directly. 

Another approach to address both equifinality specifically -- and calibration skill more generally -- is the use of idealized perfect-model experiments with synthetic observations. These provide a controlled setting to test and compare calibration frameworks based on their ability to recover known parameters. This is a standard approach for any kind of statistical algorithm development, but has so far only been implemented for very idealized model like the Lorenz model \cite{cleary_calibrate_2021}, not GCMs. \par

\subsection{Single Column Models}

Single Column Models (SCMs) offer a promising framework for evaluating and optimizing parameter calibration techniques, with much lower computational cost. SCMs remove some of the complexities of GCMs and focus on a single vertical column of the atmosphere at a specific location by isolating key physical processes \cite{gettelman_single_2019}. SCMs also contain most of the parameterizations that give rise to significant uncertainty in full GCMs, such as those related to convection, could microphysics, and radiative transfer \cite{jess_statistical_2011, bogenschutz_unified_2012, gettelman_single_2019,  neggers_attributing_2015}.  This setup thus provides a simpler yet dynamically consistent framework for analyzing critical physical mechanisms, and allows for more focused testing, debuging, and improving parameter calibration techniques.  

Since SCMs share the same foundational physical and dynamic properties as full GCMs, insights and parameter values obtained from SCM experiments can be transferred to GCM development and calibration \cite{brient_how_2012, zhang_cgils_2013}. With SCMs capable of producing meaningful results within minutes \cite{brient_how_2012}, their integration with an ML emulator and Bayesian HMC may provide a highly efficient framework for pre-training parameter sets for full GCMs and conducting analyses of observational sufficiency\par

An additional advantage of SCM frameworks is their ability to directly leverage detailed observational datasets from field campaigns and Intensive Observation Periods (IOPs). While general calibration approach rely primarily on satellite observations, which provide broad spatial coverage but relatively limited vertical resolution, field campaign datasets offer high-frequency measurements of atmospheric thermodynamic and microphysical profiles across many vertical levels. These vertically resolved observations are particularly valuable for evaluating parameterized processes such as convection and cloud microphysics, which strongly influence the vertical structure of the atmosphere. By using these high-vertical-resolution measurements as constraints, SCM-based calibration frameworks might more effectively diagnose model biases and constrain parameters controlling vertical distributions of clouds, temperature, and moisture - the dominant sources of uncertainty. 

Despite the advantages of SCMs, they do have limitations that must be considered. For example, while they simulate some of the most uncertain processes like convection and microphysics, they do not include other important and uncertain parameterizations, like those related to boundary layer turbulence. Keeping these limitations in mind, we think that SCMs have been underutilized as a steppingstone in developing better model calibrations. In addition to using SCMs to improve tuning algorithms, they can also serve as a first pre-tuning step, where they will be used to obtain a first guess of relevant parameters before tuning is attempted on the full GCM. This will hopefully reduce the number of iterations that then need to be done with the full GCM. \par

\subsection{Our Approach}

In this study, we develop a perfect-model framework based on the Single Column Atmospheric Model (SCAM) and the PPE $+$ emulation approach. We use the framework to show the drawbacks of traditional observation matching calibration and to evaluate Bayesian methods, including HMC with GP emulation.  The combination of a perfect-model framework  with a computationally efficient SCM allows us to identify sensitive parameters, quantify uncertainty, evaluate optimal calibration targets, and diagnose sensitivity to observational record lengths. The remainder of this paper is organized as follows: Section 2 describes the SCM configuration, methodology, and perfect model experimental design. Section 3 presents the results and discussion, including observation matching, equifinality, and parameter recovery using HMC. Section 4 summarizes the key findings and outlines opportunities for extending SCM-guided calibration to multi-location and multi-process studies.

\section{Models and Methods}
\subsection{Single Column Atmospheric Model (SCAM)}
We primarily use the Single Column Atmosphere Model (SCAM), a one-dimensional configuration of the Community Atmosphere Model version 6 (CAM6), specifically designed to isolate and analyze vertical atmospheric processes at a single location \cite{gettelman_single_2019}. SCAM manages vertical advection in the column by combining the full range of physics parameterizations and an advection dynamics module from CAM6. The fully interactive column radiation code and interfaces for cloud and aerosol interactions with radiation are included in SCAM. Additionally, SCAM makes use of CAM6's complete Modal Aerosol Model~\cite<MAM,>{liu_toward_2012}.\par

We choose SCAM over other SCMs for this study because it provides pre-configured forcing files for a wide range of IOP cases associated with major field campaigns \cite<Table 1 of>{gettelman_single_2019}. These forcing datasets, together with their corresponding initial conditions, enable SCAM to reproduce observed atmospheric evolution with high fidelity. Moreover, SCAM is well documented for running user-generated forcing files, making it a flexible and practical tool, and ideally suited for a perfect-model experiment.\par

\subsection{Perturbed Parameter Ensemble (PPE)}
We consider a total of 31 parameters, where 11 are related to convection, 11 to microphysics, and 9 to aerosol processes. A brief description of each parameter, along with its range and default value is provided in Table~\ref{tab:range}. These parameter ranges are based on expert judgment and are described in detail in \citeA{eidhammer_extensible_2024}. The selected parameters are among the most important and sensitive, as identified in recent studies \cite{qian_parametric_2018, eidhammer_extensible_2024, yarger_autocalibration_2024}. To generate the PPE, we use Latin Hypercube Sampling \cite<LHS;>{mckay_comparison_1979}, which draws samples randomly within the specified ranges. The range of each parameter is then divided into intervals equal to the number of samples, and each sample is assigned a value from a unique interval to ensure full coverage of the parameter space. No bin is reused across samples for any given parameter. Using this method, we generate multiple distinct parameter sets (PPEs), in addition to the SCAM (CAM6) default.

\begin{table}
\caption{\textit{A Description of the Parameters and their Ranges.} }
\centering
\resizebox{\textwidth}{!}{%
\begin{tabular}{c l l c c c l}
\hline
\textbf{Physics Scheme} & \textbf{Parameter Name} & \textbf{Description} & \textbf{Default} & \textbf{Max} & \textbf{Min} & \textbf{Unit} \\
\hline
\multirow{11}{*}{Microphysics (11)} & micro\_mg\_accre\_enhan\_fact & Accretion enhancing factor & 1.0 & 10.0 & 0.1 & \\
 & micro\_mg\_autocon\_fact & Autoconversion factor & 0.01 & 0.2 & 0.005 & \\
 & micro\_mg\_autocon\_lwp\_exp &LWP exponent & 2.47 & 3.30 & 2.1 & \\
 & micro\_mg\_autocon\_nd\_exp & Autoconversion exponent& -1.1 & -0.8 & -2.0 & \\
 & micro\_mg\_berg\_eff\_factor & Bergeron efficiency factor& 1.0 & 1.0 & 0.1 & \\
 & micro\_mg\_dcs &  Autoconversion size threshold ice–snow & 500e-6 & 1000e-6 & 50e-6 & m \\
 & micro\_mg\_effi\_factor & Scale effective radius for optics calculation & 1.0 & 2.0 & 0.1 & \\
 & micro\_mg\_homog\_size & Homogeneous freezing ice particle size& 25e-6 & 200e-6 & 10e-6 & m \\
 & micro\_mg\_iaccr\_factor & Scaling ice and snow accretion & 1.0 & 1.0 & 0.2 & \\
 & micro\_mg\_max\_nicons & Maximum allowed ice number concentration & 100e6 & 10000e6 & 1e5 & kg$^{-1}$ \\
 & micro\_mg\_vtrmi\_factor & Ice fall speed scaling& 1.0 & 5.0 & 0.2 & ms$^{-1}$ \\
\hline
\multirow{9}{*}{Aerosol (9)} & microp\_aero\_npccn\_scale &Scale activated liquid number& 1.0 & 3.0 & 0.33 & \\
 & microp\_aero\_wsub\_min &Min subgrid velocity for liquid activation& 0.2 & 0.5 & 0 & ms$^{-1}$ \\
 & microp\_aero\_wsub\_scale &Subgrid velocity for liquid activation scaling& 1.0 & 5.0 & 0.1 & \\
 & microp\_aero\_wsubi\_min &Min subgrid velocity for ice activation& 0.001 & 0.2 & 0 & ms$^{-1}$ \\
 & microp\_aero\_wsubi\_scale &Subgrid velocity for ice activation scaling& 1.0 & 5.0 & 0.1 & \\
 & dust\_emis\_fact &Dust emission scaling factor& 0.7 & 1.0 & 0.1 & \\
 & seasalt\_emis\_scale &Sea salt emission scaling factor& 1.0 & 2.5 & 0.5 & \\
 & sol\_factb\_interstitial &Below-cloud scavenging of interstitial modal aerosols& 0.1 & 1.0 & 0.1 & \\
 & sol\_factic\_interstitial &In-cloud scavenging of interstitial modal aerosols& 0.4 & 1.0 & 0.1 & \\
\hline
\multirow{11}{*}{Convection (11)} & cldfrc\_dp1 &Parameter for deep convection cloud fraction& 0.1 & 0.25 & 0.05 & \\
 & cldfrc\_dp2 &Parameter for deep convection cloud fraction& 500 & 1000 & 100.0 & \\
 & zmconv\_c0\_lnd &Convective autoconversion over land& 0.0075 & 0.1 & 0.002 & m$^{-1}$ \\
 & zmconv\_c0\_ocn &Convective autoconversion over ocean& 0.3 & 0.1 & 0.02 & m$^{-1}$ \\
 & zmconv\_capelmt &Triggering threshold for ZM convection& 70 & 350 & 35.0 & Jkg$^{-1}$ \\
 & zmconv\_dmpdz &Entrainment parameter& -1.0e-3 & -2.0e-4 & -2e-3 & m$^{-1}$ \\
 & zmconv\_ke &Convective evaporation efficiency& 5.0e-6 & 1.0e-5 & 1.0e-6 & (kgm$^{-2}$s$^{-1}$)$^{0.5}$s$^{-1}$ \\
 & zmconv\_ke\_lnd &Convective evaporation efficiency over land& 1.0e-6 & 1.0e-5 & 1.0e-5 & (kgm$^{-2}$s$^{-1}$)$^{0.5}$s$^{-1}$ \\
 & zmconv\_momcd &Efficiency of pressure term in ZM downdraft CMT& 0.7 & 1.0 & 0 & \\
 & zmconv\_num\_cin &Allowed number of negative buoyancy crossings& 1.0 & 5.0 & 1.0 & \\
 & zmconv\_tiedke\_add &Convective parcel temperature perturbation& 0.5 & 2.0 & 0 & K \\
\hline
\end{tabular}
}
\label{tab:range}
\vspace{-30pt}
\end{table}

\subsection{Gaussian Process (GP) Emulation and Bayesian Inference}
We use the open-source Earth System Emulator (ESEm), which provides a comprehensive framework for mimicking Earth system model simulations and evaluating a wide range of models and outputs \cite{watson-parris_model_2021}. This tool supports a variety of regression approaches including GP, random forest, and neural networks.
Although SCAM simulations are relatively inexpensive to run, we employ a GP emulator to efficiently explore the high-dimensional parameter space and generate continuous probabilistic predictions. The GP emulator provides probabilistic predictions, $p(Y|\theta)$, of model outputs, $Y$, for parameter combinations, $\theta$,  outside of the LHS sample, without running additional SCAM simulations.  This approach enables rapid assessment of thousands of parameter combinations, and thus facilitates robust probabilistic calibration and uncertainty quantification which would be computationally intensive if relying solely on SCAM runs.\par

In addition, we employ the Hamiltonian Monte Carlo (HMC) algorithm within the ESEm to efficiently sample the posterior distribution of the model parameters. The ESEm HMC algorithm is implemented in TensorFlow Probability. HMC generates parameter proposals using gradients of the log-posterior through Hamiltonian dynamics with leapfrog integration, followed by a Metropolis acceptance step to ensure sampling from the correct posterior distribution. This gradient-based approach enables efficient exploration of the high-dimensional parameter space and typically converges faster than conventional Metropolis–Hastings algorithms \cite{neal2011mcmc}. In the ESEm implementation, the Gaussian Process (GP) emulator provides differentiable predictions of SCAM outputs, allowing automatic computation of the log-posterior gradients required by HMC. This framework enables efficient estimation of parameter posterior distributions and facilitates recovery of the true parameters. \par 

The posterior distribution of the parameters $\theta$ conditioned on the observations $Y_0$ is given by Bayes’ theorem,
\begin{equation}
p({\theta} | {Y}_0) \propto p({Y}_0 | {\theta}) \times p({\theta}),
\end{equation}
where $p(\theta)$ represents the prior distribution of the parameters and $p(Y_0|\theta)$ is the likelihood of the observations given the parameters. In practice, the likelihood $p({Y}_0|\theta)$ is approximated by a normal distribution centered on the emulator mean $\mu_E (\theta)$ with total variance $\sigma_t^2$, which accounts for multiple sources of uncertainty:
\begin{equation}
p({Y}_0|\theta) \approx
\frac{1}{\sigma_t\sqrt{2\pi}}
\exp\left[-\frac{1}{2}\left(\frac{{Y}_0 - \mu_E (\theta)}{\sigma_t}\right)^2\right],
\quad
\sigma_t = \sqrt{\sigma_E^2 + \sigma_Y^2 + \sigma_R^2 + \sigma_S^2}.
\label{eq:likelihood}
\end{equation}
Here, $\sigma_E^2$ represents emulator uncertainty, $\sigma_Y^2$ observational uncertainty, $\sigma_R^2$ representational uncertainty, and $\sigma_S^2$ structural model uncertainty. Since this study is conducted as a perfect-model experiment, the observational ($\sigma_Y^2$), structural ($\sigma_S^2$), and representational ($\sigma_R^2$) uncertainties are all assumed to be zero. In ESEm, the emulator uncertainty, $\sigma_Y^2$ is internally estimated from the predictive variance of the emulator.

\vspace{-0.5cm}
\begin{table}[H]
\centering
\setlength{\abovecaptionskip}{2pt}
\caption{\textit{Key variables and corresponding units used for calibrating the SCAM PPE.}}
\label{tab:variables}
\resizebox{0.5\textwidth}{!}{%
\small
\begin{tabular}{l l l}
\hline
\textbf{Variable} & \textbf{ID} & \textbf{Unit} \\
\hline
Cloud Fraction & CLOUD & - \\
Liquid Water Path & TGCLDLWP & g/kg \\
Relative Humidity & RH & - \\
Residual Top-of-model Energy Balance & RESTOM  &  W/m\textsuperscript{2} \\
Short Wave Cloud Forcing & SWCF & W/m\textsuperscript{2} \\
Long Wave Cloud Forcing & LWCF & W/m\textsuperscript{2} \\
Temperature & T & K \\
\hline
\end{tabular}%
}
\end{table}

\subsection{Perfect Model Experiment}

We conduct a perfect model experiment using SCAM at the same location as the Southern Great Plains (SGP) observatory of the Department of Energy's Atmospheric Radiation Measurement (ARM) program. We force SCAM with user-generated forcing from CAM6, following \citeA{gettelman_single_2019}, thus  providing a controlled setting for assessing parameter sensitivity and evaluating the performance of the proposed approaches. The default parameter values serve as a ``synthetic truth" that will be the target of our calibration efforts, while the SCAM output using these default value serves as ``synthetic observations”. We refer to this setup as a perfect-model experiment, since both the PPEs and the synthetic observations are generated from the same model using the same configuration, just different parameter values. The existence of a ``true" set of parameters allows for objective assessment of calibration frameworks. \par 

We integrate the model over a full year (months 1–12) at the SGP. However, for this initial study, the analysis is restricted to the period from April 1 (day 91) to July 31 (day 212). This four-month period corresponds to the late spring and early summer season when the SGP frequently experiences deep convection, making it particularly suitable for evaluating convective and cloud-related processes.\par

We then run SCAM with a 100-member PPE (hereafter, 100PPE) and a 500-member PPE (hereafter, 500PPE), ensuring that all ensemble members are integrated under identical forcing and boundary conditions at SGP. Our objective is to retrieve the true parameters using synthetic observations for variables listed in Table~\ref{tab:variables}. These have been chosen to match both actual observations available at the DOE ARM SGP site, as well as standard variables used in past PPE-based calibration approaches. The 100PPE and 500PPE simulations are designed to evaluate both point estimation and probabilistic calibration, examine equifinality, and assess whether increasing ensemble size improves constraints on parameter uncertainties. \par

The GP emulator is trained using outputs from both ensembles (100PPE and 500PPE) and each version employed to generate 10,000-member ensembles. These GP emulators are subsequently coupled with the HMC algorithm to perform efficient Bayesian inference of the parameters. This combined GP–HMC framework enables rapid sampling from the posterior distribution while significantly reducing computational cost compared to direct SCAM integrations. \par

\begin{figure}[H]
\centering
\includegraphics[width=\textwidth]{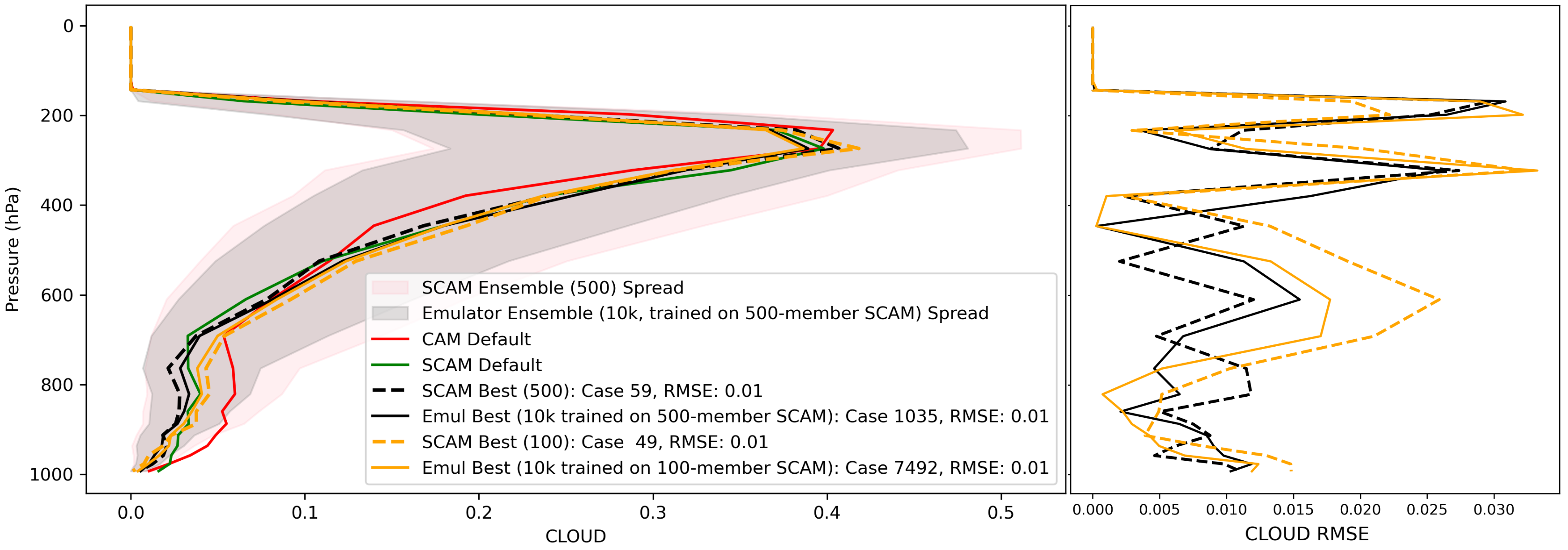}
\caption{Time Mean Vertical Profile of cloud fraction (left panel) and RMSE from SCAM default (right panel). The pink and gray shading denotes the 5th to 95th percentile spread of the 500-member SCAM PPE and the GP-emulated 10,000-member ensemble trained on those SCAM outputs respectively. The red line shows the CAM CLOUD profile, and the green line corresponds to the synthetic observation (SO). The black dashed line marks the SCAM member that best matches the synthetic observations within the 500-member SCAM ensemble, and the black solid line shows the emulator’s best-match member. Whereas the yellow dashed and solid lines are same as black lines but for 100-member SCAM PPE and the GP-emulated 10,000-member ensemble trained on those SCAM outputs.}
\label{fig:cld_vert}
\end{figure}

\section{Results and Discussions}

\subsection{GP emulation results}
\begin{figure}[H]
\centering
\includegraphics[width=\textwidth]{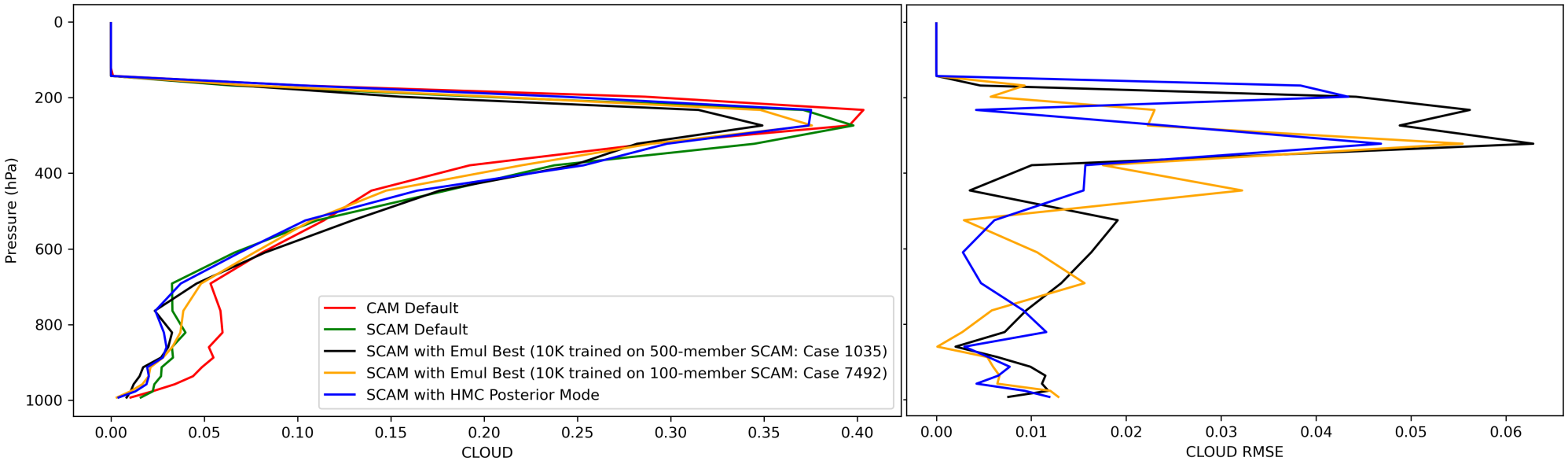}
\caption{Time-mean vertical profiles of cloud fraction (left panel) and RMSE from the true state (right panel), based on SCAM simulations using the best parameter set inferred from emulation and the HMC posterior mode.}
\label{fig:cld_vert_all_best}
\end{figure}

\begin{figure}[H]
\centering
\includegraphics[width=\textwidth]{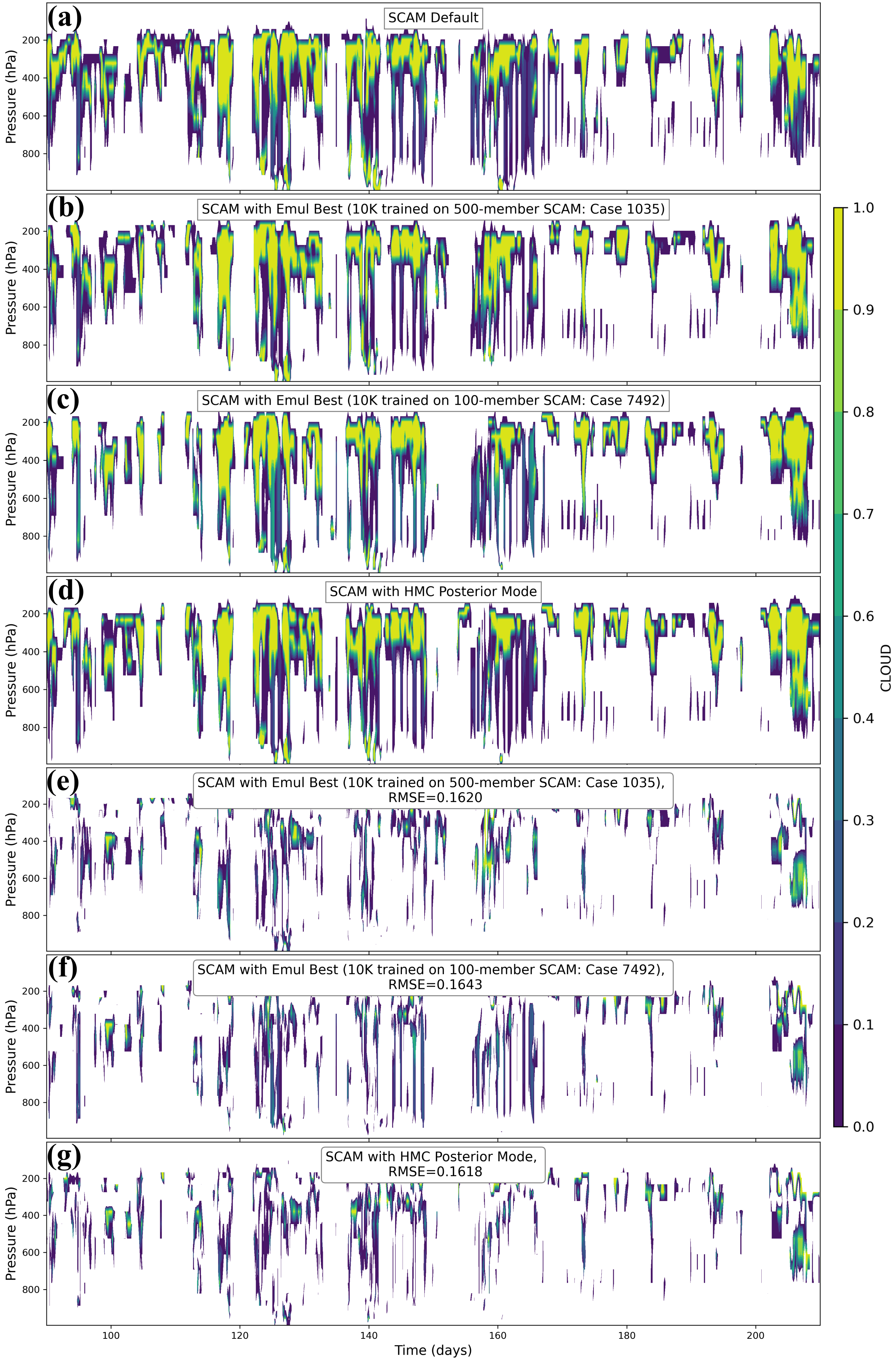}
\caption{Temporal evolution of cloud fraction: (a) SCAM default, (b–d) SCAM simulations with best case parameter sets, and (e–g) corresponding RMSE relative to the default. }
\label{fig:cld_temporal}
\end{figure}

We begin by looking at the skill of the GP emulator, when trained on observations of time-resolved vertical cloud fraction (CLOUD).  Figure~\ref{fig:cld_vert} and~\ref{fig:cld_vert_all_best} then shows the the time mean vertical profile of CLOUD, while Figures ~\ref{fig:cld_temporal} show time resolved CLOUD.  The CAM (Figure~\ref{fig:cld_vert}, red line) and SCAM default (Figure \ref{fig:cld_vert}, green line, SO) closely match in the mid- to upper troposphere, with previously documented deviations in the lower troposphere \cite{gettelman_single_2019}, demonstrating that SCAM effectively reproduces the full CAM simulation. In addition, the strong overlap and similar structure of the pink and gray shading indicate that the GP emulator accurately captures the ensemble characteristics of SCAM. 

We first consider four ``best" model cases. The SCAM Best (100) and SCAM Best (500) members are the members of 100PPE and 500PPE with the lowest RMSE relative to the synthetic CLOUD observations, while Emul Best (100) Emul and Best (500) are the members of the two 10,000 member emulator ensembles with the lowest RMSE between the emulated CLOUD output and synthetic CLOUD observations. All four best cases (Figure \ref{fig:cld_vert}, black and yellow lines) closely align with the synthetic observations (green line) across most levels, indicating optimal PPE sampling and strong emulation fidelity. The emulator is also able to reproduce the broad structure and magnitude of different other variables (supplementary Figures \ref{fig:rehum_vert} and \ref{fig: temp_lwp}), demonstrating that the GP emulator effectively captures the SCAM responses.

To evaluate emulator uncertainty, we perform SCAM simulations using the ``emulated best" parameter sets and the GP-HMC posterior mode parameter set. Passing these parameter sets back to SCAM results in CLOUD profiles that continue to closely match the the synthetic observations Figure~\ref{fig:cld_vert_all_best}), confirming relatively low emulator error. We further examine the temporal evolution of cloud profiles by comparing these simulations with SCAM default configuration (SO) (Figure~\ref{fig:cld_temporal}). Consistent with the vertical profile results, the HMC posterior mode produces the lowest RMSE across all best cases (Figure~\ref{fig:cld_temporal}g).

\begin{figure}[H]
\centering
\includegraphics[width=\textwidth]{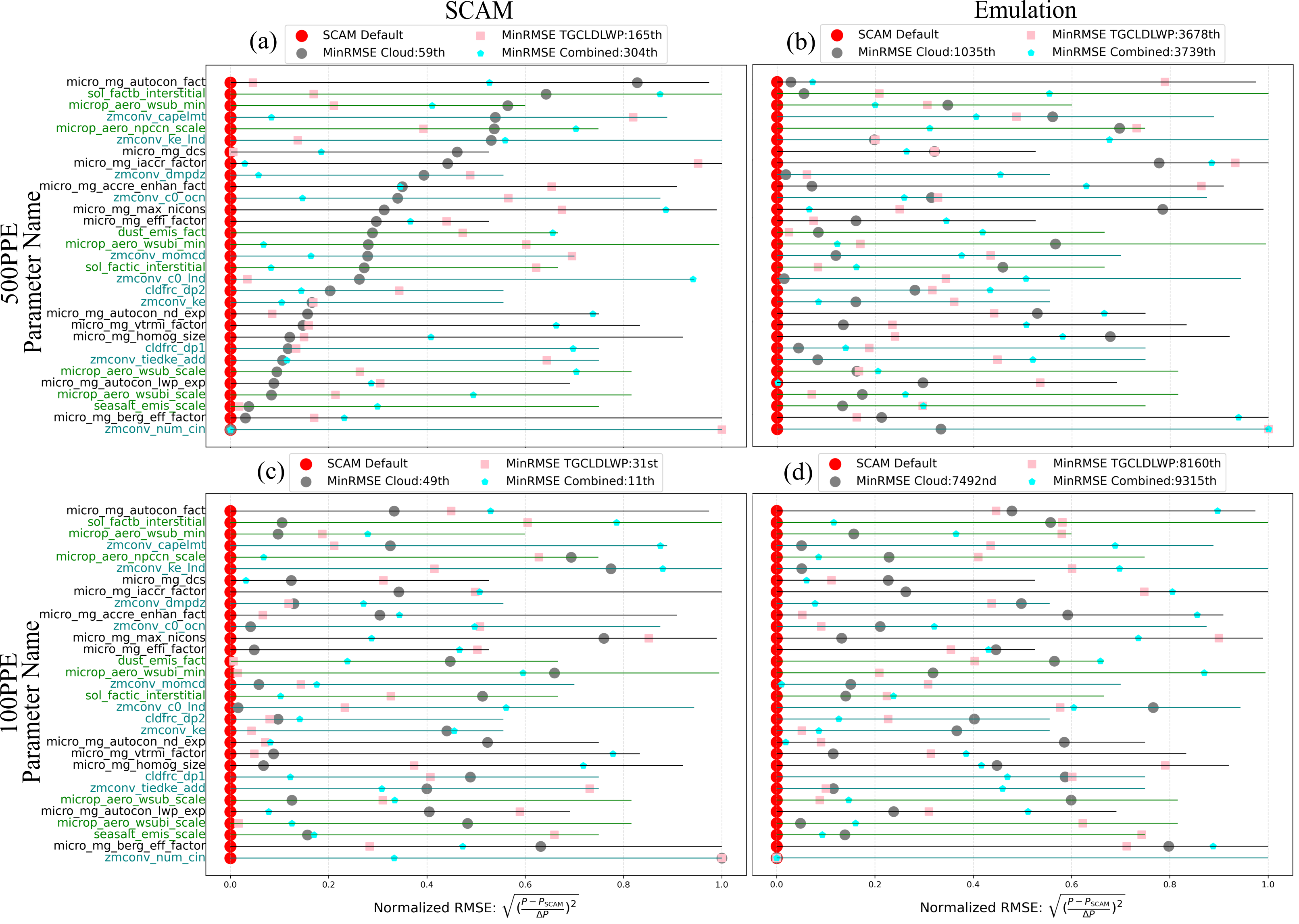}
\caption{Parameter Error Space: Parameter Name vs. Normalized RMSE. Parameter names are color-coded: black for microphysics, green for aerosols, and teal for convection. Panels (a) and (b) correspond to the 500-member SCAM PPE and its emulator predictions (10K-member), panels (c) and (d) show the same for the 100-member PPE and its emulator predictions. The horizontal lines of each panel represents the range of RMSE for each parameter across the ensemble members. The parameter names sorted by best-case for CLOUD of panel (a) (gray circles) RMSE from lowest (bottom) to highest (top). Red circles indicate the true parameter RMSE (zero) from the synthetic observations simulation (SCAM default). Pink rectangles show the best-case parameters for TGCLDLWP, and cyan pentagons represent the best-case parameters for the normalized combination of all variables listed in Table~\ref{tab:variables}. The combined parameter RMSE is computed by first normalizing each variable in SCAM and SO, then evaluating the best case of the normalized multi-variable fields.}
\label{fig:error}
\end{figure}
\vspace{-1cm}

\subsection{Equifinality in Parameter Recovery by Observation Fitting}
The ``best cases" simulation and the default simulation that produced the synthetic observations  used identical forcing and atmospheric conditions at the same location. The output from the best cases closely aligns with the synthetic observations and exhibit extremely low root mean square error. Therefore, the parameter values involved in these best cases should be close to the default parameter value in the synthetic observations simulation. 

Despite the close agreement in vertical CLOUD (RMSE~0.01 for all four best cases), the parameter values associated with these best-match profiles differ substantially. For example, the best-match profile in the 500PPE ensemble corresponds to 59th case, while the 100PPE ensemble corresponds to 49th case, and the parameters of these two cases are scattered all over their possible (Figure~\ref{fig:error}a,c). It is also evident from Figure~\ref{fig:error}a that the  corresponding parameter values retrieved by matching the best model version to synthetic observations deviate largely from their true values in most parameters, with only a few exceptions, such as $zmconv\_num\_cin$ when calibrating to CLOUD and combined variables, where $micro\_mg\_dcs$ and $seasalt\_emis\_scale$ when calibrating to TGCLDLWP. In fact, most of the parameters exhibit substantial scatter for all other variables (Figure~\ref{fig: error_all}). We do the same comparison for GP emulation in Figure~\ref{fig:error}b and find the same pattern of random and widespread scatter. Here, however, a few different parameters such as $zmconv\_c0\_lnd$, and $zmconv\_dmpdz$ for CLOUD, $dust\_emis\_fact$ for TGCLDLWP, $micro\_mg\_autocon\_lwp\_exp$ for combined are constrained correctly. Additionally, we include the SCAM 100PPE and the corresponding GP emulation predictions (Figure~\ref{fig:error}c,d) and observe a similarly widespread and seemingly random distribution of best-case parameter sets, which remain far from the true values for most parameter. Moreover, in each case, 2–3 random parameters are correctly retrieved.\par

To assess whether incorporating additional data improves parameter calibration, we extend the analysis to include both temporal evolution and vertical structure, rather than relying solely on time-mean vertical profiles. Despite the increased information content, we find a similarly broad distribution of best-case parameter sets, with different parameter combinations yielding comparable fits to the observations (Supplementary Figures~\ref{fig:error_fulldata} and~\ref{fig: error_all_fulldata}). This indicates that even when accounting for full temporal and vertical variability, the parameter space remains poorly constrained, highlighting the persistence of equifinality in observation-based calibration. \par

\begin{figure}[H]
\centering
\includegraphics[width=\textwidth]{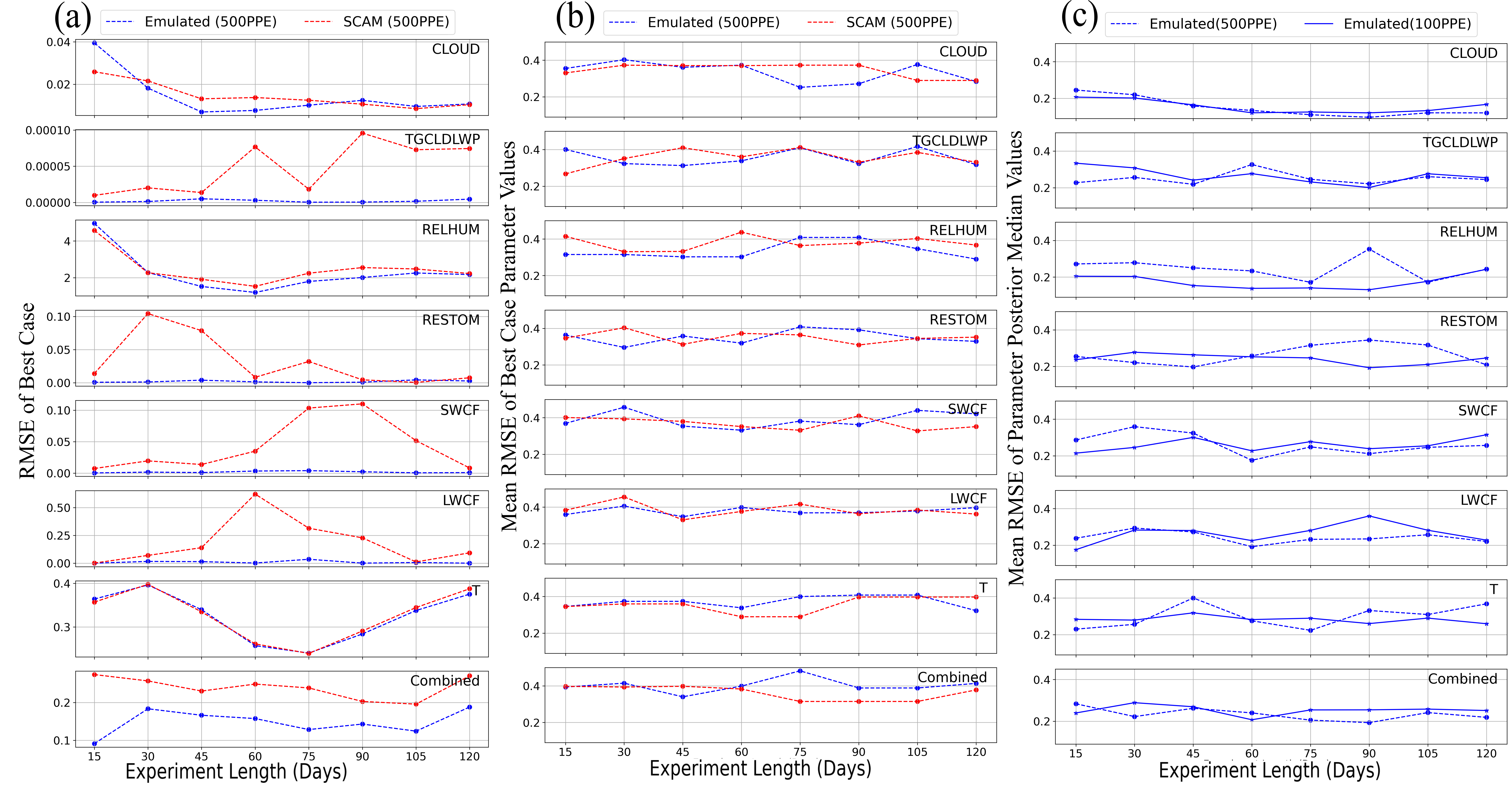}
\caption{RMSE of variables listed in table~\ref{tab:variables} and parameters as a function of experiment length (15–120 days). Colors are consistent across panels: red = SCAM , blue = Emulated ; dashed lines indicate results from the 500PPE experiments, solid lines correspond to the 100PPE experiment. (a) The RMSE of each individual variable and the combined (top row) relative to the synthetic observations. The combined (multi-variable) RMSE is computed by first normalizing each variable in SCAM and synthetic observations, then evaluating the RMSE of the normalized fields. (b) Normalized mean RMSE of best case parameter values derived from observation matching (figure~\ref{fig:error}), (c) normalized mean RMSE of nine most sensitive parameters posterior-median values obtained from GP-HMC.}
\label{fig:line}
\end{figure}

These behaviors are clear manifestation of the equifinality problem in earth system models, where distinct parameter combinations produce nearly identical model outputs. Increasing the ensemble size does not resolve this, as the observation-matching process selects a different parameter combination in each of the best cases (Figure~\ref{fig:error} and \ref{fig: error_all}). Equifinality is best highlighted by the fact that the emulator is able to find an ensemble member with near zero RMSE in the target variable when calibrating to radiation fields such as RESTOM, LWCF, and SWCF (Figure~\ref{fig:line}a), indicating that the best ensemble member matches the synthetic observations almost perfectly. However, the normalized mean RMSE for the best-case parameter set is relatively high (around 0.4; Figure~\ref{fig:line}b), suggesting that the inferred optimal parameters deviate significantly from the true parameter values. 

We also analyze the impact of observational record length,  by using synthetic observational records starting from April 1 with 15-day increments through July 31. This allows us to test whether longer sampling periods improve parameter recovery. However, even with extended sampling, naive point-estimation fails to recover the true or near-true parameter values. Although 60-day experiments generally minimize RMSE for most variables (Figure~\ref{fig:line}a), this does not translate into accurate parameter estimation (Figure~\ref{fig:line}b). For example, in the 60-day experiment, the RMSE of the best-performing case approaches near-zero values. However, the normalized mean RMSE of the corresponding parameter values ranges between 0.3 and 0.4, indicating that the parameter set producing the lowest RMSE for cloud-radiation fields remain substantially different from the true parameter values. This strongly suggests that point estimates of parameters obtained by minimizing the error of different fields does not necessarily lead to recovery of the true parameters. 
    \par

\vspace{-0.5cm}
\begin{figure}[H]
\centering
\includegraphics[width=\textwidth]{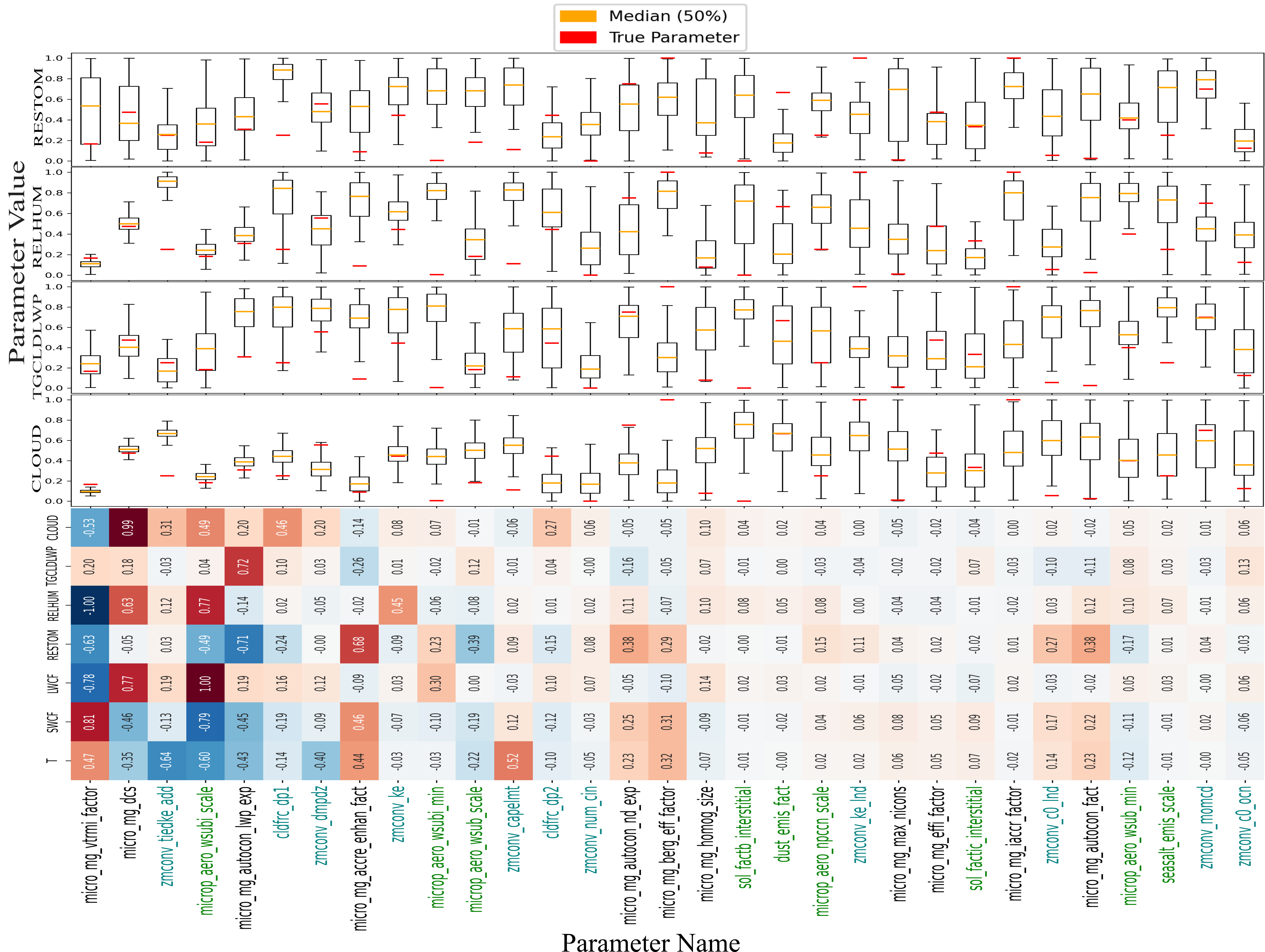}
\caption{Normalized posterior distributions parameters (top four panels), shown as inter-quartile ranges (IQR, 25th–75th percentiles), and corresponding parameter sensitivity analysis (bottom panel). The posterior distributions panels are derived from GP-HMC trained on 500 PPE using 60-day experiment, for the variables CLOUD, TGCLDLWP, RELHUM, and RESTOM. The parameter name are sorted by arranging the box height of posterior for CLOUD , with the smallest IQR on the left and the largest on the right. Yellow horizontal lines denote posterior medians, and red horizontal lines indicate the true parameter values.The bottom panel is the the regression-derived normalized parameter sensitivities, where reds represent positive sensitivity and blues represent negative sensitivity; the numerical sensitivity values are annotated in each cell. Parameter name colors follow the same microphysics–aerosol–convection grouping described in Figure~\ref{fig:error}. }
\label{fig:sen_box}
\end{figure}
\vspace{-1cm}

\subsection{Overcoming Equifinality with GP-HMC Based Parameter Estimation}
Observation-matching fails to recover the true parameters due to pervasive equifinality, even under a perfect-model framework. By leveraging 500PPE and 100PPE trained GP-HMC, we reformulate this ill-posed inverse problem into a statistically constrained one that explicitly incorporates prior distribution, parameter covariances, and the full shape of the likelihood surface. In GP–HMC posterior analysis, the four-month analysis period is used without applying temporal averaging so that the full temporal variability is retained and a larger number of data points are available to constrain the parameter estimation. Here, we consider a 60-day experiment length for the posterior analysis because the mean RMSE of GP-HMC parameter posterior-median  for most sensitive parameters (Figure \ref{fig:line}c) shows that 60 days yields the lowest RMSE for most cloud-radiation fields. \par

The HMC driven posterior distributions (Figure~\ref{fig:sen_box} and \ref{fig:sen_box_100en}) demonstrate that it successfully constrains the parameters that exert strong sensitivity on the model variables. For instance, $micro\_mg\_vtrmi\_factor$ one of the most influential parameters across all SCAM output fields listed on bottom panel of Figure~\ref{fig:sen_box} exhibits a very narrow inter-quartile range (IQR) for CLOUD, TGCLDLWP, and RELHUM (three middle panels of Figure~\ref{fig:sen_box}). In addition, the posterior median closely aligned with the true value. A similar pattern is observed among the first nine parameters, from $micro\_mg\_vtrmi\_factor$ to $zmconv\_ke$, which are also identified as the most sensitive parameters. Therefore, GP-HMC can effectively collapse the uncertainty and recover values close to the truth for highly influential parameters. We repeat this analysis using GP-HMC trained on a 100PPE and find that the same subset of parameters is consistently retrieved (Figure~\ref{fig:sen_box_100en}).However, the accuracy of the recovered parameters is substantially improved when using the 500PPE GP-HMC. The larger ensemble training provides a more informative approximation of the parameter–response relationships, resulting in narrower posterior distributions and more reliable retrieval of the true parameters. We are able to perform the 500PPE GP-HMC training easily because SCAM is computationally inexpensive, enabling us to run large ensembles that would be prohibitive or extremely costly with a full GCM. \par

However, within this strongly sensitive parameter region, radiation field particularly RESTOM (similarly, SWCF and LWCF, not shown) exhibit relatively weak posterior constraint. This suggests that despite being sensitive, the radiative fields do not provide sufficient gradient information for certain parameters. HMC also does not recover parameters that show weak sensitivity across all variables, but this limitation is expected since these parameters have little influence on the model outputs and thus are inherently unidentifiable. Another feature emerges, for example the parameter $zmconv\_tiedke\_add$ for CLOUD, and RELHUM are well constrained but remain far from the true parameter values. This may be due to interactions with other parameters, as they are varied simultaneously and exhibit strong interdependencies. Overall, these results show that HMC provides a substantial improvement over naive observation matching by reliably constraining the most sensitive parameters except for those primarily tied to the radiation fields. \par

Another notable feature of the GP–HMC parameter posterior distributions for the nine sensitive parameters is that the associated RMSE values fall within the range of 0.1–0.2, which is substantially smaller than the RMSE range obtained from the observation-fitting parameter sets (Figure \ref{fig:line}b, c). This behavior is consistent across all other variables, where the RMSE ranges are also significantly lower than those from the observation-based parameter fitting. This reduction in RMSE suggests that the GP–HMC framework yields more reliable parameter estimates.

\section{Conclusions}

Our study demonstrates a perfect model framework for calibrating the Single Column Atmospheric Model (SCAM). We find that  point estimates of parameters -- finding one parameter set that best matches observations -- is not a reliable method for identifying true parameter values. Even when using synthetic observations from a perfect-model setup, observation matching fails  to recover the underlying parameters. This result highlights the limitations of relying solely on best-match or profile-based comparison to observational data for parameter tuning and emphasizes the need for methods that account for uncertainty and parameter sensitivity more rigorously.

In contrast, adopting a Bayesian probabilistic framework using GP-HMC to derive posterior distributions proves to more effective. This approach is able to identify the most sensitive parameters and quantify the associated uncertainties, providing a robust probabilistic parameter estimation. Another important feature of GP-HMC is its ability to consistently recover the same set of parameters from both the 500PPE and 100PPE ensemble training. This demonstrates that the probabilistic approach functions reliably, in contrast to RMSE-based observation matching which typically retrieves only a few random parameters in each instance. While the posteriors are mostly consistent, they do occasionally provide wrong, overconfident estimates for some parameters (e.g. $zmconv\_tiedke\_add$). Also, as expected, a large number of parameters remain unidentifiable given the available synthetic observations. 

The perfect model framework is useful in identifying optimal tuning targets. Our results show that tuning to certain variables, like vertically resolved cloud fraction, offers significantly more calibration skill than other variables such as radiation fields, relative humidity, or liquid water path. While not addressed, here, perfect model frameworks can be useful in creating cost functions that optimally weight different observable variables. The results also show how parameter calibration skill depends on record length, with the best results obtained when calibrating to CLOUD, using at least 60 days. An important use of perfect model frameworks could be to help inform IOPs, by asking what measurements, season, location, and deployment length would provide optimal data for constraining cloud microphysics, aerosol, or convection processes in climate models. Finally, perfect model tests could be used to identify if other variables (in other locatoins and seasons) could help constrain the parameters that are not constrainable with the observations and location considered here. 

Despite these promising results, there are several limitations to the current study. The analysis was conducted for a single location at SGP and used only synthetic observations, which limits the generalizability of the conclusions. Future work should extend this approach to other locations and incorporate real-world observational datasets from ARM ground-based campaigns, and aircraft data in addition to satellites observations could provide stronger and more realistic constraints on model parameters. Furthermore, while SCAM offers a computationally efficient and practical framework for constraining sensitive parameters, its greatest utility lies in serving as a “training step” for full three-dimensional climate models. In future work, we aim to apply the SCAM-identified constrained parameters to guide and pre-condition the tuning of the full GCM, hoping to substantially reduce the overall computational cost of model tuning. Additionally, while we focus exclusively on HMC for posterior estimation, other ensemble-based approaches such as EnKF or its variants offer promising alternatives for parameter tuning. SCMs are very amenable to EnKF methods, and could provide complementary insights into parameter sensitivities.

\section*{Open Research Section}
All data and code to reproduce the results shown are available at 

https://zenodo.org/records/19380115

\section*{Conflict of Interest declaration}
The authors declare that there are no conflicts of interest for this manuscript.

\acknowledgments
CP and PP were supported by the Department of Energy (DOE) Award \# DE-SC0022110 through the Regional and Global Model Analysis (RGMA) program.

%
%

\bibliography{ agusample }

%
%
%
%
%

\clearpage
\section*{Supplementary}

\setcounter{figure}{0}
\renewcommand{\thefigure}{S\arabic{figure}}

\begin{figure}[H]
\centering
\includegraphics[width=\textwidth]{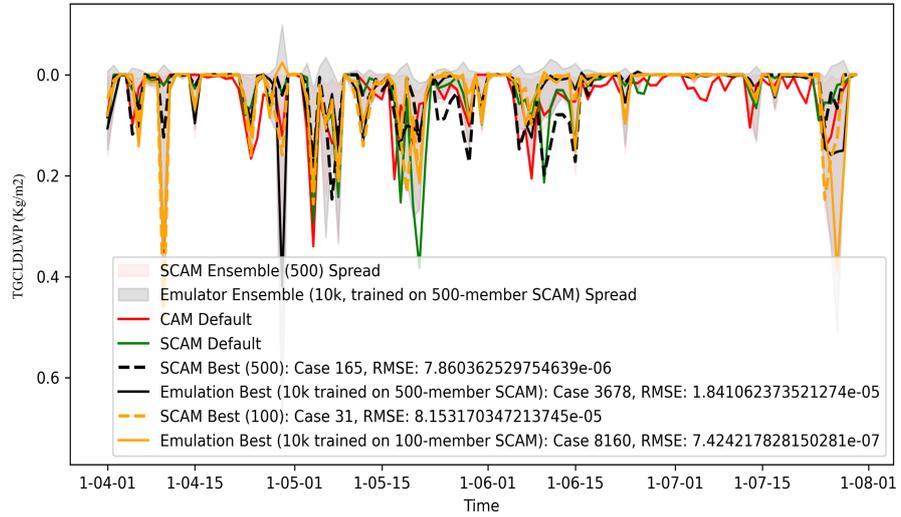}
\caption{Same as  figure~\ref{fig:cld_vert} but for Temporal profile of liquid water path (TGCLDLWP).}
\label{fig: temp_lwp}
\end{figure}

\begin{figure}[H]
\centering
\includegraphics[width=\textwidth]{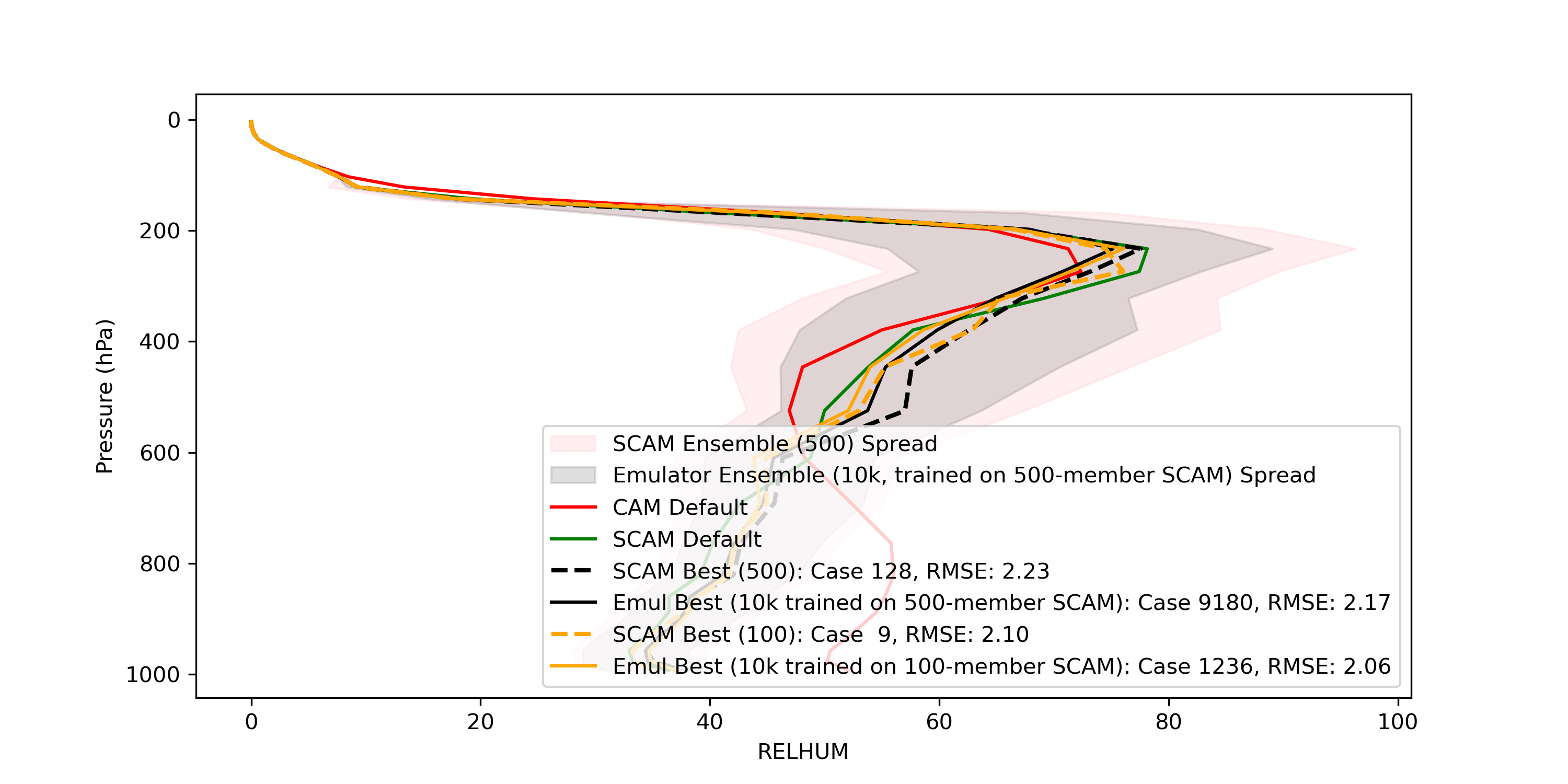}
\caption{Same as figure~\ref{fig:cld_vert} but for Vertical profile of Relative Humidity (RELHUM).}
\label{fig:rehum_vert}
\end{figure}

\begin{figure}[H]
\centering
\includegraphics[width=\textwidth]{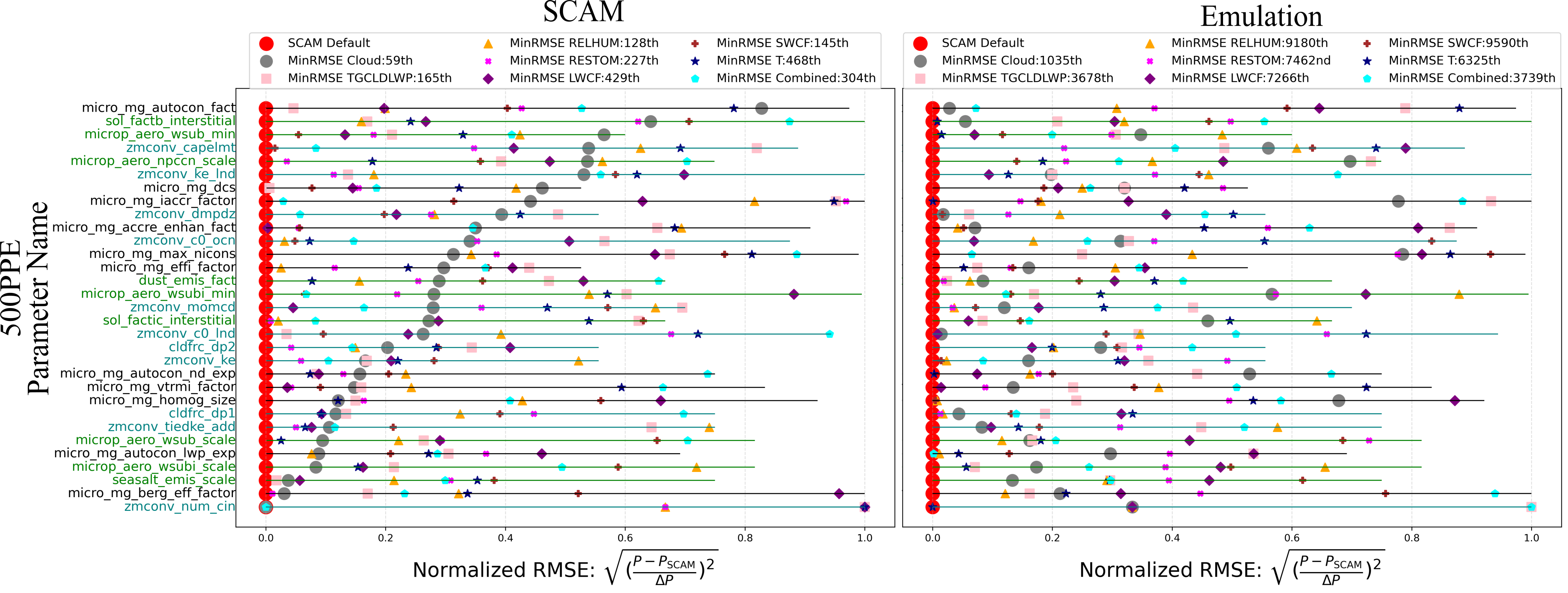}
\caption{Same as  figure~\ref{fig:error}a,b but for all variables.}
\label{fig: error_all}
\end{figure}

\begin{figure}[H]
\centering
\includegraphics[width=\textwidth]{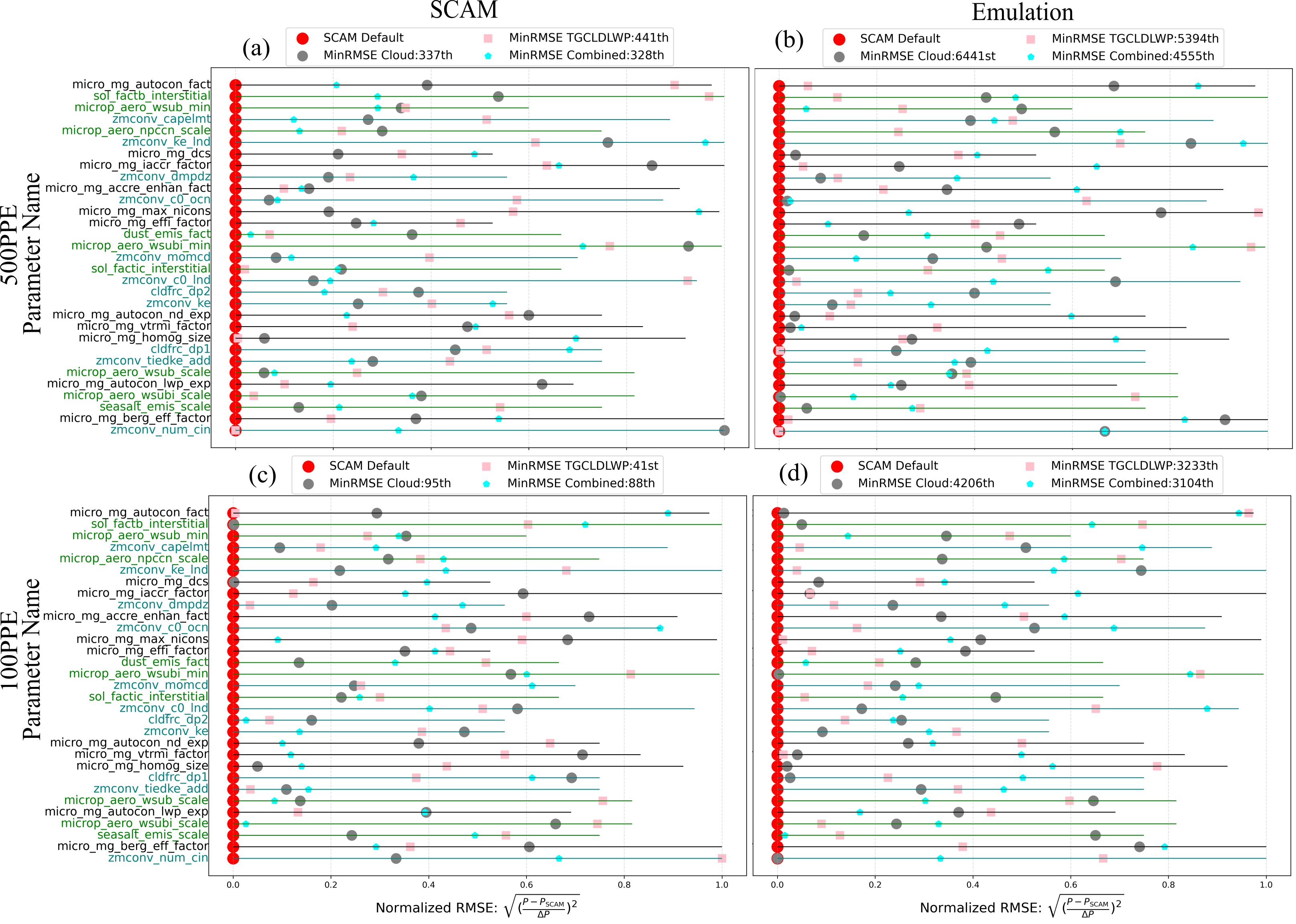}
\caption{Same as Figure~\ref{fig:error}, but computed using the full dataset, incorporating both temporal variability and vertical structure.}
\label{fig:error_fulldata}
\end{figure}

\begin{figure}[H]
\centering
\includegraphics[width=\textwidth]{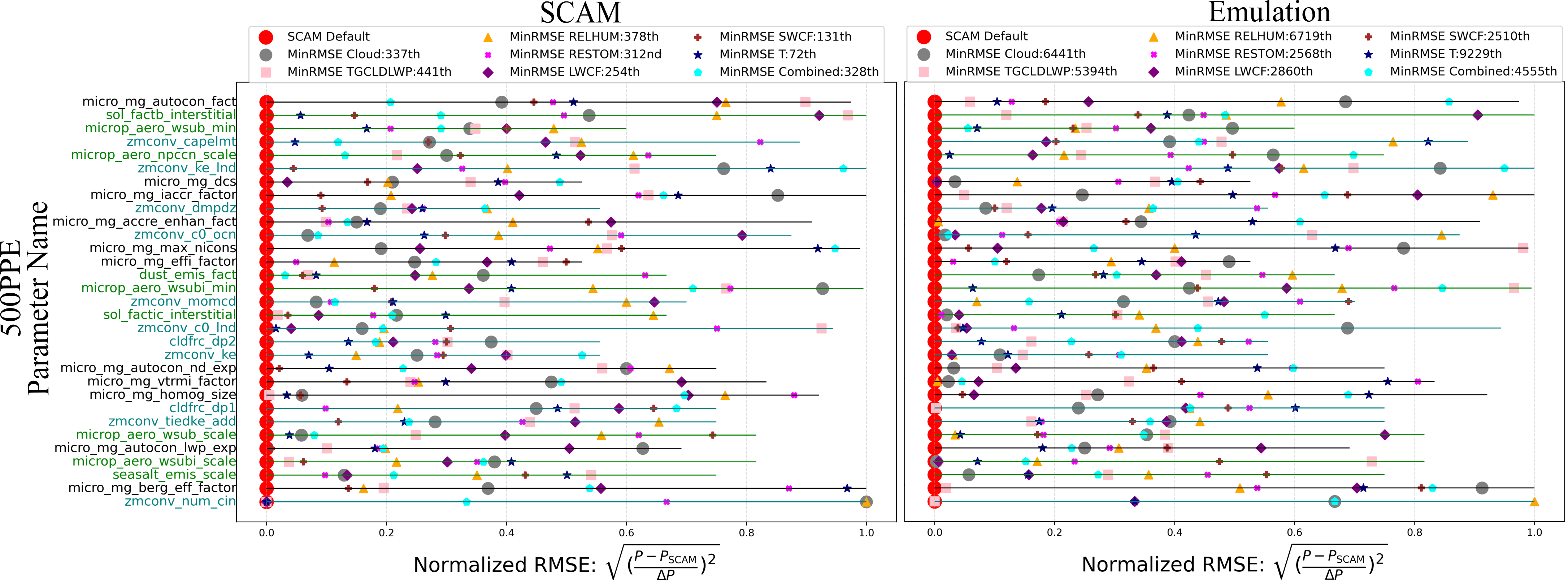}
\caption{Same as Figure~\ref{fig: error_all}, but computed using the full dataset, incorporating both temporal variability and vertical structure.}
\label{fig: error_all_fulldata}
\end{figure}

\begin{figure}[H]
\centering
\includegraphics[width=\textwidth]{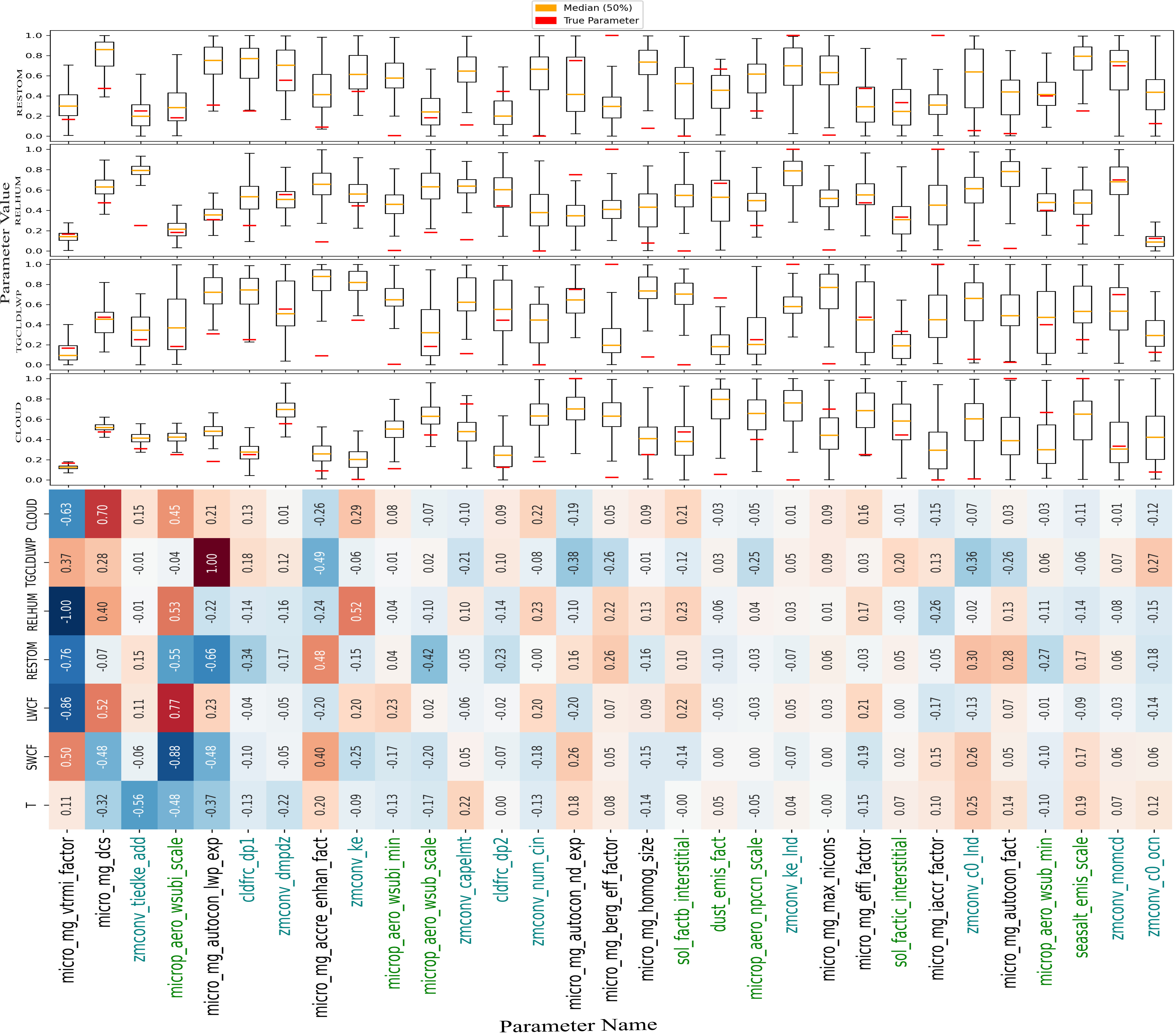}
\caption{Same as Figure~\ref{fig:sen_box} but for GP-HMC trained on 100PPE.}
\label{fig:sen_box_100en}
\end{figure}

\end{document}